\batchmode
\makeatletter
\def\input@path{{D:/Australia/thz/meeting/JCS/}}
\makeatother
\documentclass[twocolumn,journal]{IEEEtran}
\usepackage[T1]{fontenc}
\usepackage[latin9]{inputenc}
\usepackage{array}
\usepackage{float}
\usepackage{textcomp}
\usepackage{mathrsfs}
\usepackage{amsmath}
\usepackage{amsthm}
\usepackage{amssymb}
\usepackage{stackrel}
\usepackage{graphicx}
\usepackage[pdftex,unicode=true,
 bookmarks=true,bookmarksnumbered=true,bookmarksopen=true,bookmarksopenlevel=1,
 breaklinks=false,pdfborder={0 0 0},pdfborderstyle={},backref=false,colorlinks=false]
 {hyperref}
\hypersetup{pdftitle={Your Title},
 pdfauthor={Your Name},
 pdfpagelayout=OneColumn, pdfnewwindow=true, pdfstartview=XYZ, plainpages=false}

\makeatletter

\newcommand{\lyxmathsym}[1]{\ifmmode\begingroup\def\b@ld{bold}
  \text{\ifx\math@version\b@ld\bfseries\fi#1}\endgroup\else#1\fi}

\providecommand{\tabularnewline}{\\}
\floatstyle{ruled}
\newfloat{algorithm}{tbp}{loa}
\providecommand{\algorithmname}{Algorithm}
\floatname{algorithm}{\protect\algorithmname}

\let\oldforeign@language\foreign@language
\DeclareRobustCommand{\foreign@language}[1]{%
  \lowercase{\oldforeign@language{#1}}}
\theoremstyle{plain}
\newtheorem{thm}{\protect\theoremname}

\usepackage[caption=false,font=footnotesize]{subfig}

\@ifundefined{showcaptionsetup}{}{%
 \PassOptionsToPackage{caption=false}{subfig}}
\usepackage{subfig}
\makeatother

\providecommand{\theoremname}{Theorem}

\begin{document}
\title{Hybrid 3D Beamforming Relying on Sensor-Based Training and Channel
Estimation for Reconfigurable Intelligent Surface Aided TeraHertz
MIMO systems }
\author{Xufang~Wang,~\IEEEmembership{Member,~IEEE,} Zihuai~Lin,~\emph{Senior
Member,~IEEE},~Feng~Lin,~\IEEEmembership{Member,~IEEE,} and~Lajos~Hanzo,~\IEEEmembership{Life Fellow,~IEEE}\thanks{Xufang Wang is with the Key Laboratory of Optoelectronic Science and
Technology for Medicine of Ministry of Education, Fujian Normal University,
Fuzhou, China (e-mail: fzwxf@fjnu.edu.cn). }\thanks{Zihuai Lin is with the School of Electrical and Information Engineering,
University of Sydney, Sydney, NSW 2006, Australia (e-mail: zihuai.lin@sydney.edu.au).}\thanks{Feng~Lin is with Kongtronics Institute of Science and Technology
(XiaMen) Co., Ltd., Xia Men, China (e-mail: helloyou@189.cn).}\thanks{Lajos Hanzo is with the University of Southampton, SO17 1BJ Southampton,
U.K. (e-mail: lh@ecs.soton.ac.uk).}}
\markboth{}{Xufang Wang \MakeLowercase{\emph{et al.}}: Hybrid 3D Beamforming
Relying on Sensor-based Training and Channel Estimation for Reconfigurable
Intelligent Surfaces Aided Terahertz MIMO systems}
\maketitle
\begin{abstract}
\emph{Terahertz} (THz) systems have the benefit of high bandwidth
and hence are capable of supporting ultra-high data rates, albeit
at the cost of high pathloss. Hence they tend to harness high-gain
beamforming. Therefore a novel hybrid 3D beamformer relying on sophisticated
sensor-based beam training and channel estimation is proposed for
\emph{Reconfigurable Intelligent Surface} (RIS) aided THz MIMO systems.
A so-called array-of-subarray based THz BS architecture is adopted
and the corresponding sub-RIS structure is proposed. The BS, RIS and
receiver antenna arrays of the users are all \emph{uniform planar
arrays} (UPAs). The \emph{Ultra-wideband} (UWB) sensors are integrated
into the RIS and the user location information obtained by the UWB
sensors is exploited for channel estimation and beamforming. Furthermore,
the novel concept of a \emph{Precise Beamforming Algorithm} (PBA)
is proposed, which further improves the beamforming accuracy by circumventing
the performance limitations imposed by positioning errors. Moreover,
the conditions of maintaining the orthogonality of the RIS-aided THz
channel are derived in support of spatial multiplexing. The closed-form
expressions of the near-field and far-field path-loss are also derived.
Our simulation results show that the proposed scheme accurately estimates
the RIS-aided THz channel and the spectral efficiency is much improved,
despite its low complexity. This makes our solution eminently suitable
for delay-sensitive applications.
\end{abstract}


\IEEEpeerreviewmaketitle{}

\section{Introduction}

\IEEEPARstart{T}{}he data rate requirements of wireless communications
have increased rapidly with the explosive growth of mobile devices
and seamless multimedia applications. Hence, the bandwidth available
in the sub-6 \emph{Gigahertz} (GHz) and mmWave bands becomes tight,
but as a remedy, the \emph{Terahertz} (THz) range of (0.1-10 THz)
may be explored for 6G development \cite{elayan_terahertz_2020}\cite{wang_performance_2020}.
Given its short wavelength, a large number of antennas can be packed
into a compact transceiver. Graphene based plasmonic massive MIMO
nano-antenna arrays have been developed in \cite{akyildiz_realizing_2016}
for the THz band, relying on beamforming based angular multiplexing
\cite{sarieddeen_terahertz-band_2019}. However, traditional systems
using a dedicated RF chain for each antenna are no longer practical
due to the unacceptable hardware costs. In \cite{lin_energy-efficient_2016},
a hybrid beamforming system associated with an array-of-subarray structure
is proposed for THz communications, in which the number of RF chains
is much smaller than the number of antennas. 

In addition, although THz communication has the advantage of increasing
the bandwidth by orders of magnitude, it inherently suffers from limited
coverage range due to the diffusion loss caused by high frequencies,
the absorption of molecules in the atmospheric medium and the higher
probability of \emph{line-of-sight} (LOS) blockage \cite{ma_joint_2020}.
A feasible way forward is to combine the massive MIMO transciever
with a \emph{reconfigurable intelligent surface} (RIS), whose phase
shift can be controlled by a low-complexity programmable PIN diode
\cite{pan_intelligent_2020} in support of 6G systems \cite{pan_reconfigurable_2021}.
It can improve the propagation conditions by introducing controllable
scattering to obtain beneficial beamforming gains and mitigate the
interference \cite{wu_intelligent_2019}. This is achieved by increasing
the freedom of transceiver design and network optimization by intelligently
ameliorating the wireless propagation environments \cite{di_renzo_smart_2020}. 

Despite the fact that the RIS optimization of most MISO systems can
be directly transformed into quadratic programming under quadratic
constraints, there is a paucity of literature on the design of RIS
assisted MIMO systems \cite{zhang_capacity_2020,ye_joint_2020}. Hence,
we conceive a MIMO-aided and RIS assisted THz system. Additionally,
most of the open literature assumes that the channel state information
(CSI) of the RIS is perfectly available \cite{wu_intelligent_2019}\cite{huang_reconfigurable_2019}\cite{guo_weighted_2020}.
However, since all elements of the RIS are passive, it cannot send,
receive or process any pilot signal for channel estimation. Moreover,
RISs usually contain hundreds of elements, so the dimension of the
estimated channel is much larger than that of traditional systems,
hence resulting in an excessive pilot overhead. Therefore, traditional
solutions cannot be directly applied, and channel estimation is a
key challenge \cite{wei_channel_2021}. Therefore, sophisticated schemes
have been proposed for the channel estimation of RIS-aided MISO systems
\cite{ma_joint_2020} \cite{elbir_deep_2020} \cite{he_cascaded_2020}
\cite{liu_deep_2020}, where the receiver is equipped with a single
antenna. However, these schemes cannot be readily combined with the
above-mentioned channel estimation schemes in massive MIMO systems.
A cooperative beam training scheme is developed in \cite{ning_terahertz_2021}
to facilitate the estimation of the concatenated twin-hop BS-RIS-UE
channel. However, they assumed having no obstacles between the BS
and users, which may not always be the case. In \cite{hu_location_2020},
the location information obtained by GPS is used to assist RIS systems,
but the altitude error of GPS is quite high, about twice as high as
that on the ground, which is unacceptable for channel estimation in
3D beamforming. Moreover, GPS often fails in obstructed open areas
or indoors. By comparison, UWB wireless positioning can reach centimeter
level accuracy \cite{sharma_ir-uwb_2018}. 

On the other hand, given its compelling benefits, Integrated Sensing
and Communication (ISAC) \cite{liu_joint_2020} have attracted substantial
research attention. The ISAC concept also influences the\emph{ sixth
generation} (6G) network design \cite{wild_joint_2021}. In this context,
a number of ISAC schemes have been proposed. For example, a unified
and reconfigurable multi-functional receiver is presented in \cite{moghaddasi_multifunctional_2016}
for data fusion in radar sensing and radio communication. As a further
development, a new ISAC system is given in \cite{zhang_multibeam_2019}
for multi-beamforming based mobile communication relying on a\emph{
time division duplex} (TDD) framework. A new beamforming technique
is proposed in \cite{liu_mu-mimo_2018} subject to specific \emph{signal-to-noise
ratio} (SINR) constraints. They also proposed a mmWave Massive MIMO
ISAC system operating in the face of interference \cite{liu_joint_2020}.
An ISAC antenna composed of a sensing subarray and a communication
subarray is presented for UAVs \cite{chen_performance_2020}.

In this paper, a novel hybrid 3D beamforming with sensor-based beam
training and channel estimation scheme is proposed for RIS assisted
THz MIMO systems. Explicitly, an array-of-subarray based THz BS architecture
is adopted and the corresponding sub-RIS structure is proposed. The
RIS can be installed both on building walls and facades for supporting
3D passive beamforming. The BS, RIS and the receiver antenna arrays
are all \emph{uniform planar arrays} (UPAs). The UWB sensors are integrated
into the RIS and the user location information obtained by UWB sensors
is used for channel estimation and beamforming. A so-called \emph{Precise
Beamforming Algorithm} (PBA) is proposed, which is capable of improving
the beamforming accuracy by eliminating the deleterious effects of
positioning errors. Our simulation results show that the proposed
scheme accurately estimates the RIS-aided THz channel. As a benefit
of the RIS, the spectral efficiency becomes much higher than that
of the system operating without RISs. 

Against the above backdrop, our main contributions are:

1) We propose the hybrid 3D beamforming BS array-of-subarray and the
corresponding sub-RIS structure for RIS-aided THz MIMO systems. Furthermore,
we derive the conditions of orthogonality for the RIS-aided THz channel
in support of spatial multiplexing;

2) We conceive a UWB sensor-based channel estimator for RIS-aided
THz MIMO systems. The PBA concept based on user location information
is proposed for improving the beamforming accuracy. Thus, the system
performance will no longer be constrained by the positioning errors
of the sensors, yet compared to other beam training schemes our proposed
scheme has the lowest complexity and search time, rendering it eminently
suitable for users having time-varying positions or delay-sensitive
applications;

3) We derive the closed-form expressions for both the near-field and
far-field path-loss of RIS-aided THz channels. Both the near-field
and far-field scenarios demonstrate the efficiency of our proposed
scheme. 

We also included Table I for boldly contrasting the novelty of our
paper to the state-of-the-art. 

\begin{table*}[tbh]
\caption{Our novel contributions contrasted to the state-of-the art}

\centering{}%
\begin{tabular}{|>{\raggedright}m{30mm}|>{\centering}m{9mm}|>{\centering}m{8mm}|>{\centering}m{8mm}|>{\centering}m{8mm}|>{\centering}m{8mm}|>{\centering}m{8mm}|>{\centering}m{8mm}|>{\centering}m{8mm}|>{\centering}m{8mm}|>{\centering}m{8mm}|>{\centering}m{8mm}|}
\hline 
 &
our paper &
\centering{}\cite{lin_adaptive_2015}-2015 &
\centering{}\cite{wu_intelligent_2019}-2019 &
\centering{}\cite{ma_joint_2020}-2019 &
\centering{}\cite{pan_intelligent_2020}-2019 &
\centering{}\cite{zhang_capacity_2020}-2019 &
\centering{}\cite{hu_location_2020}-2020 &
\centering{}\cite{tang_wireless_2021}-2021 &
\centering{}\cite{sarieddeen_terahertz-band_2019}-2021 &
\centering{}\cite{guo_weighted_2020}-2021 &
\centering{}\cite{ning_terahertz_2021}-2021\tabularnewline
\hline 
\hline 
\raggedright{}\textbf{Sensor-based channel estimation} &
\centering{}$\surd$ &
 &
 &
 &
\centering{} &
\centering{} &
\centering{} &
 &
 &
\centering{} &
\tabularnewline
\hline 
\raggedright{}\textbf{Hybrid 3D beamforming} &
\centering{}$\surd$ &
\centering{} &
\centering{} &
\centering{} &
\centering{} &
\centering{} &
\centering{} &
\centering{} &
\centering{} &
\centering{} &
\centering{}\tabularnewline
\hline 
\raggedright{}RIS aided systems &
\centering{}$\surd$ &
\centering{} &
\centering{}$\surd$ &
\centering{}$\surd$ &
\centering{}$\surd$ &
\centering{}$\surd$ &
\centering{}$\surd$ &
\centering{}$\surd$ &
\centering{} &
\centering{}$\surd$ &
\centering{}$\surd$\tabularnewline
\hline 
\raggedright{}THz Communications &
\centering{}$\surd$ &
\centering{}$\surd$ &
\centering{} &
\centering{}$\surd$ &
\centering{} &
\centering{} &
\centering{} &
\centering{} &
\centering{}$\surd$ &
\centering{} &
\centering{}$\surd$\tabularnewline
\hline 
\raggedright{}BS array-of-subarray  &
\centering{}$\surd$ &
$\surd$ &
 &
 &
 &
 &
 &
 &
\centering{}$\surd$ &
 &
\tabularnewline
\hline 
\raggedright{}\textbf{sub-RIS structure} &
\centering{}$\surd$ &
 &
 &
 &
 &
 &
 &
 &
 &
 &
\tabularnewline
\hline 
\raggedright{}\textbf{Near-field and far-field path-loss of RIS-aided
THz channels} &
\centering{}$\surd$ &
\centering{} &
\centering{} &
\centering{} &
\centering{} &
\centering{} &
\centering{} &
 &
\centering{} &
\centering{} &
\centering{}\tabularnewline
\hline 
\raggedright{}\textbf{Precise beam forming} &
\centering{}$\surd$ &
\centering{} &
\centering{} &
\centering{} &
\centering{} &
\centering{} &
\centering{} &
\centering{} &
\centering{} &
\centering{} &
\centering{}\tabularnewline
\hline 
\raggedright{}MIMO systems &
$\surd$ &
 &
 &
\centering{}$\surd$ &
\centering{}$\surd$ &
\centering{}$\surd$ &
 &
 &
\centering{}$\surd$ &
 &
\centering{}$\surd$\tabularnewline
\hline 
\end{tabular}
\end{table*}

The remainder of the paper is organized as follows. In Section II,
we describe both the system model and channel models. In Section III,
we explore the channel conditions to be satisfied for high-integrity
spatial multiplexing over the RIS-aided THz channels. In Section IV,
the closed-form expressions of the path-loss of the RIS-aided near-field
and far-field beamforming channels are derived. In Section V, the
proposed sensor based channel estimation scheme is presented. Our
performance analysis is discussed in Section VI, followed by our simulation
results in Section VII. Finally, we conclude in Section VIII. 

\emph{Notation: }Boldface lower case and upper case letters are used
for column vectors and matrices, respectively. The superscripts $(\cdot)^{*}$
, $(\cdot)^{T}$, $(\cdot)^{H}$, and $(\cdot)^{-1}$ stand for the
conjugate, transpose, conjugate-transpose, and matrix inverse, respectively.
The Euclidean norm, absolute value, Hadamard product are denoted by
$||\cdot||$, $|\cdot|$ and $\odot$ respectively. In addition, $\mathbb{E}\left\{ \cdot\right\} $
is the expectation operator. For a matrix $\mathbf{A}$, $[\mathbf{A}]_{mn}$
denotes its entry in the \emph{m}-th row and \emph{n}-th column, while
for a vector $\mathbf{a}$, $[\mathbf{a}]_{m}$ denotes the \emph{m}-th
entry of it. Furthermore, $j$ in $e^{j\theta}$ denotes the imaginary
unit, while $\mathbf{I}$ is the identity matrix. 

\section{System Model and Channel Models}

In this section, we introduce the system model and the channel models
of 3D hybrid beamforming designed for RIS-assisted THz MIMO systems,
including the direct BS-to-user path and the BS-RIS-user path.

\begin{figure*}[tbh]
\begin{centering}
\includegraphics[scale=0.6]{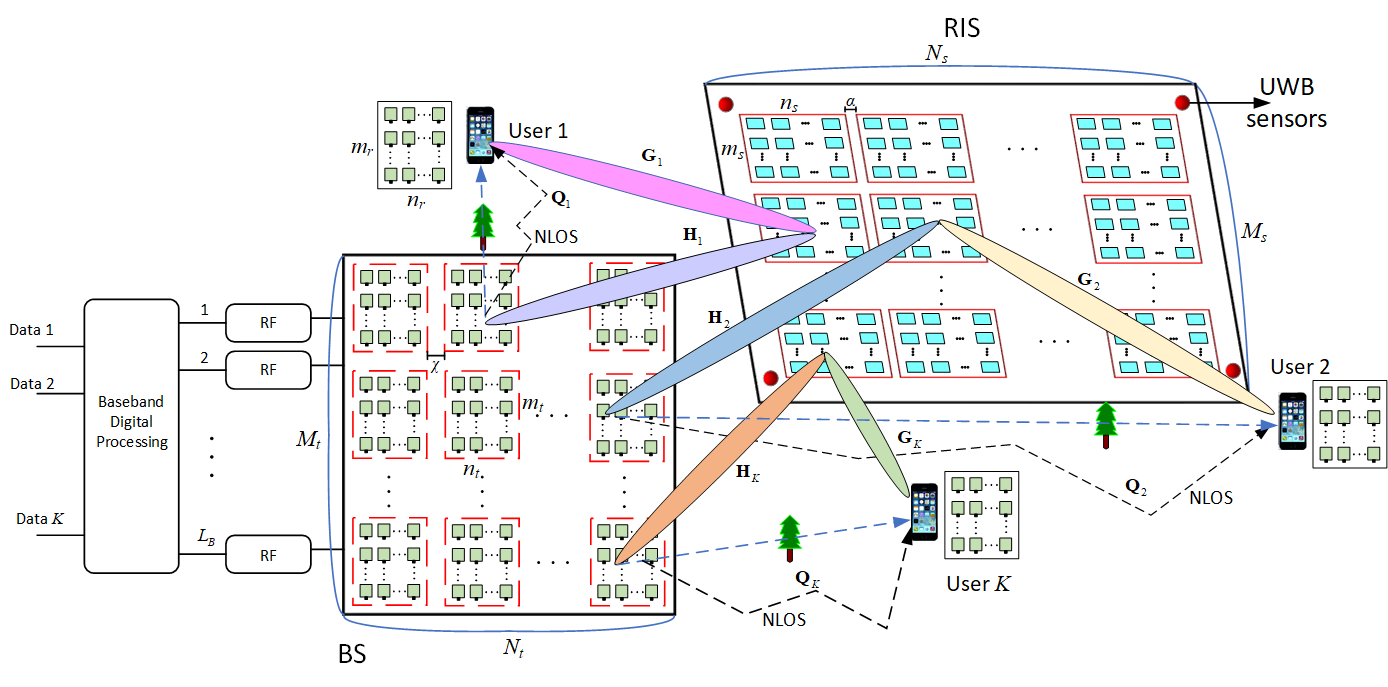}
\par\end{centering}
\caption{\label{fig:System-model-of}System model of 3D hybrid beamforming
and sensor-based channel estimation for RIS-aided THz MIMO systems.}
\end{figure*}

The system model we adopted is shown in Fig. \ref{fig:System-model-of}.
The THz transceivers have array-of-subarrays of graphene-based plasmonic
nano-antennas \cite{akyildiz_realizing_2016}. The BS transmitter
(TX) having $L_{B}=M_{t}\times N_{t}$ subarrays supports \emph{$K$}
users either with or without RISs. The \emph{i}th subarray of the
BS is a UPA having $m_{t,i}\times n_{t,i}$ antenna units. For simplicity
and without loss of generality, we let $m_{t,i}=m_{t}$, $n_{t,i}=n_{t},$
$i=1,...,L_{B}$. Note that $L_{B}$ is also the number of RF chains,
since each BS subarray is controlled by a dedicated RF chain. Due
to limited processing power, there is only a single subarray baseband
and RF chain consisting of $m_{r,k}\times n_{r,k}$ tightly-packed
elements at the \emph{k}th user. For simplicity and without loss of
generality, we let $m_{r,k}=m_{r}$, $n_{r,k}=n_{r},$ $k=1,...,K$.
The number $L_{B}$ of antenna subarrays is assumed to be higher than
\emph{$K$} for attaining high gains. UPAs are promising for THz communications
both in BS and UE terminals, since they can accommodate more antenna
elements by a two-dimensional subarray for 3D beamforming.

In THz communications, the link from the BS to the user may be blocked.
Hence we assume that the line-of-sight (LOS) link from the BS to each
user is indeed blocked, and a RIS is applied for improving the link
spanning from the BS to the user by reflecting the signal. The RIS
consists of a sub-wavelength UPA having $\mathbb{N}$ passive reflecting
elements under the control of an RIS controller. The THz channel is
highly frequency-selective, but there are several low-attenuation
windows separated by high-attenuation spectral nulls owing to molecular
absorption \cite{moldovan_and_2014}. Therefore, we can adaptively
partition the total bandwidth into numerous subbands. If the bandwidth
of the subband adopted is small enough, the channel can be regarded
as non-frequency-selective and the noise power spectral density appears
to be locally flat. In the following, we will discuss the direct path
spanning from the BS to the user and that from the BS to the user
via RIS, i.e., the BS-RIS-UE path.

(1) \textbf{\emph{The direct BS-user path}}

The received signal of user $k$ can be expressed as

\begin{equation}
\bar{y}_{k}=\mathbf{v}_{k}^{H}\mathbf{Q}_{k}\mathbf{WFs}+\mathbf{v}_{k}^{H}\mathbf{n}_{k},~k=1,...,~K,\label{eq:direct}
\end{equation}
where $\mathbf{s}=[s_{1},s_{2},...,s_{K}]^{T}$ is a $K\times1$ vector
containing the transmitted symbols of $K$ users, so that $E[\mathbf{ss}^{*}]=\frac{P_{S}}{K}\mathbf{I}_{K}$,
where $P_{s}$ denotes the total initial transmit power and the same
power is assigned to each user. In (1), $\mathbf{Q}_{k}$ is the $m_{r}n_{r}\times L_{B}m_{t}n_{t}$
THz channel matrix between the BS and user $k$. The LOS components
of the direct BS-user links are blocked by obstacles, so we assume
that the direct link channel $\mathbf{\mathbf{Q}_{\mathit{k}}}$ contains
only \emph{non-line-of-sight} (NLOS) components; $\mathbf{W}$ is
the $L_{B}m_{t}n_{t}\times L_{B}$ analog transmit beamforming matrix
representing the equal power splitters and phase shifters. For the
array-of-subarray structure of (\ref{eq:direct}), $\mathbf{W}$ is
block-diagonal structure and can be expressed as

\begin{equation}
\mathbf{W}=\left[\begin{array}{cccc}
\mathbf{w}_{1} & \mathbf{0} & \cdots & \mathbf{0}\\
\mathbf{0} & \mathbf{w}_{2} & \cdots & \mathbf{0}\\
\vdots & \vdots & \vdots & \vdots\\
\mathbf{0} & \mathbf{0} & \cdots & \mathbf{w}_{L_{B}}
\end{array}\right],
\end{equation}
where $\mathbf{w}_{l}$ is an $m_{t}n_{t}\times1$ vector with $|\mathbf{\left(\mathbf{w}_{\mathit{l}}\right)}_{a,b}|=\frac{1}{\sqrt{m_{t}n_{t}}},$
$l=1,...,L_{B}$, $a=1,...,m_{t}$, and $b=1,...,n_{t}$. Still referring
to (1), $\mathbf{F}$ is the $L_{B}\times K$ baseband digital beamforming
matrix used for interference mitigation, where we have
\begin{equation}
\mathbf{F}=\left[\mathbf{f}_{1},\mathbf{\mathbf{f}}_{2},\ldots,\mathbf{\mathbf{f}}_{K}\right],
\end{equation}
with 
\begin{equation}
||\mathbf{W}\mathbf{f}_{k}||^{2}=1;\label{eq:wfj}
\end{equation}
$\mathbf{f}_{k}$ is a $L_{B}\times1$ vector of the \emph{k}th user;
$\mathbf{v}_{k}$ is the $m_{r}n_{r}\times1$ receive analog beamforming
vector applied by user \emph{$k$} with $|\mathbf{\left(v_{\mathit{k}}\right)}_{a,b}|=\frac{1}{\sqrt{m_{r}n_{r}}}$
, $a=1,...,m_{r}$, and $b=1,...,n_{r}$; $\mathbf{n}_{k}$ is the
$m_{r}n_{r}\times1$ Gaussian noise vector at user $k$, i.e., $\mathbf{n}_{k}\sim\mathscr{\mathcal{N}}(\mathbf{0},\sigma_{k}^{2}\mathbf{I})$,
$k=1\text{,...,~\emph{K}}$. 

(2) \textbf{\emph{The RIS-aided path}}

The RIS consists of a sub-wavelength UPA having $\mathbb{N}$ passive
reflecting elements. According to the subarray structure of the BS,
we also partition the RIS into $L_{s}=M_{s}\times N_{s}$ sub-RISs.
The $i$th sub-RIS consists of $m_{s,i}\times n_{s,i}$ elements,
$i=1,...,L_{s}$. Each sub-RIS is the dual counterpart of every subarray
of the BS. Thus, we have $\mathbb{N}=\stackrel[i=1]{L_{s}}{\sum}m_{s,i}n_{s,i}$.
The adjacent elements are separated by $\varrho\geq\lambda_{spp}$,
where $\lambda_{spp}$ is the \emph{Surface Plasmon Polariton} (SPP)
wavelength \cite{sarieddeen_terahertz-band_2019}, so the mutual coupling
between the elements is neglected. For simplicity and without loss
of generality, we let $m_{s,i}=m_{s}$, $n_{s,i}=n_{s},$ $i=1,...,L_{s}$.

Let the THz channels spanning from the BS to RIS, and from the RIS
to user $k$, be denoted by $\mathbf{H}_{k}$ and $\mathbf{G}_{k}$,
respectively. The RIS reflection matrix is denoted by $\mathbf{O}$
and the reflection matrix from the $k$th sub-RIS to user $k$ is
denoted by $\mathbf{O}_{k}$. For the sub-RIS structure, $\mathbf{O}$
is an $L_{s}m_{s}n_{s}\times L_{s}m_{s}n_{s}$ block matrix

\begin{equation}
\mathbf{\mathbf{O}}=\left[\begin{array}{cccc}
\mathbf{O}_{1} & \mathbf{0} & \cdots & \mathbf{0}\\
\mathbf{0} & \mathbf{\mathbf{O}}_{2} & \cdots & \mathbf{0}\\
\vdots & \vdots & \vdots & \vdots\\
\mathbf{0} & \mathbf{0} & \cdots & \mathbf{\mathbf{O}}_{L_{s}}
\end{array}\right],
\end{equation}
where $\mathbf{O}_{k}$ is an $m_{s}n_{s}\times m_{s}n_{s}$ matrix;
$\mathbf{O}_{k}=\mathrm{diag}(\mathbf{q}_{k})$, where $\mathbf{q}_{k}=\left[e^{k\theta_{1,k}},\ldots,e^{k\theta_{i,k}},\ldots,e^{k\theta_{m_{s}n_{s},k}}\right]^{T}$,
$\theta_{i,k}$ is the phase of the $i$-th reflection element in
the \emph{k}-th sub-RIS, $i=1,...,m_{s}n_{s}$. Without loss of generality,
we assume that the \emph{k}-th sub-RIS serves the \emph{k}-th user.
So the number of serving sub-RISs satisfies $L_{s}=K$. 

The received signal of user $k$ can be expressed as

\begin{equation}
\tilde{y}_{k}=\mathbf{v}_{k}^{H}\mathbf{H}_{k}\mathbf{O}\mathbf{G}\mathbf{WFs}+\mathbf{v}_{k}^{H}\mathbf{n}_{k},\quad\forall k\in\mathcal{K}\label{eq:via ris}
\end{equation}
where $\mathbf{G}$ is the $L_{s}m_{s}n_{s}\times L_{B}m_{t}n_{t}$
element channel matrix of the line spanning from the BS to the RIS.
Furthermore, $\mathbf{H}_{k}$ is the $m_{r}n_{r}\times L_{s}m_{s}n_{s}$
channel matrix of the line emerging from the\emph{ }RIS to the user\emph{
k. }Observe that $\tilde{y}_{k}$ can also be expressed in the form
of the desired and interference terms as follows:
\begin{equation}
\tilde{y}_{k}=\mathbf{v}_{k}^{H}\mathbf{H}_{k}\mathbf{O}\mathbf{G}\mathbf{W}\mathbf{f}_{k}s_{k}+\mathbf{v}_{k}^{H}\mathbf{H}_{k}\mathbf{O}\mathbf{G}\sum_{i\neq k}^{K}\mathbf{W}\mathbf{f}_{i}s_{i}+\mathbf{v}_{k}^{H}\mathbf{n}_{k}.\label{eq:received signal}
\end{equation}

The RIS-aided THz channels are assumed to consist a direct BS-RIS
ray from LOS propagation and some indirect BS-RIS rays from NLOS propagation
due to reflection and scattering \cite{lin_adaptive_2015}. Thus,
$\mathbf{G_{\mathit{k}}}$ and $\mathbf{H_{\mathit{k}}}$ are modeled
by 

\begin{align}
\mathbf{G_{\mathit{k}}} & =\bar{\mathbf{G}}_{\mathit{k}}+\mathbf{\tilde{G}_{\mathit{k}}}\nonumber \\
 & =\sqrt{m_{t}n_{t}m_{s}n_{s}}[\beta_{1,k}^{L}\mathbf{a}_{sa,k}\left(\delta_{k},\kappa_{k}\right)\mathbf{a}_{t,k}^{H}\left(\psi_{k},\sigma_{k}\right)\nonumber \\
 & +\stackrel[i=1]{n_{NL}}{\sum}\beta_{1,k,i}^{NL}\mathbf{a}_{sa,k}\left(\delta_{k,i}^{NL},\kappa_{k,i}^{NL}\right)\mathbf{a}_{t,k}^{H}\left(\psi_{k,i}^{NL},\sigma_{k,i}^{NL}\right)],\label{eq:G}
\end{align}

\begin{align}
\mathbf{H}_{k} & =\mathbf{\bar{H}_{\mathit{k}}}+\mathbf{\tilde{H}_{\mathit{k}}}\nonumber \\
 & =\sqrt{m_{r}n_{r}m_{s}n_{s}}[\beta_{2,k}^{L}\mathbf{a}_{r,k}\left(\vartheta_{k},\phi_{k}\right)\mathbf{a}_{sd,k}^{H}\left(\varsigma_{k},\varphi_{k}\right)\nonumber \\
 & +\stackrel[i=1]{\widetilde{n}_{NL}}{\sum}\beta_{2,k,i}^{NL}\mathbf{a}_{r,k}\left(\vartheta_{k,i}^{NL},\phi_{k,i}^{NL}\right)\mathbf{a}_{sd,k}^{H}\left(\varsigma_{k,i}^{NL},\varphi_{k,i}^{NL}\right)].\label{eq:arj}
\end{align}
where $\bar{\mathbf{G}}_{\mathit{k}}$ and\emph{ $\mathbf{\bar{H}_{\mathit{k}}}$
}represent the\emph{ }LOS components, while $\mathbf{\tilde{G}_{\mathit{k}}}$\emph{
}and $\mathbf{\tilde{H}_{\mathit{k}}}$ denote the NLOS components;
$\psi_{k}$ ($\psi_{k,i}^{NL}$), $\sigma_{k}$ ($\sigma_{k,i}^{NL})$
are the azimuth and elevation AODs (\emph{angle of departure}) at
the $k$-th BS subarray, respectively; $\vartheta_{k}$ ($\vartheta_{k,i}^{NL}$),
$\phi_{k}$ ($\phi_{k,i}^{NL}$) are the azimuth and elevation AOAs
(\emph{angle of arrival}) at the $k$-th user, respectively; $\delta_{k}$
($\delta_{k,i}^{NL}$), $\kappa_{k}$$(\kappa_{k,i}^{NL})$ are the
azimuth and elevation AOAs from the $k$-th BS subarray to the $k$-th
sub-RIS, respectively; $\varsigma_{k}$($\varsigma_{k,i}^{NL}$),
$\varphi_{k}$ ($\varphi_{k,i}^{NL}$) are the azimuth and elevation
AODs from the $k$-th sub-RIS to the $k$-th user, respectively; $n_{NL}$
and $\widetilde{n}_{NL}$ are the numbers of NLOS components; $\beta_{1,k}^{L}$
and $\beta_{2,k}^{L}$ denote the corresponding THz LOS complex gains
given by

\begin{align}
|\beta_{g,k}^{L}|^{2} & =\xi_{g,k}^{L}\left(d_{g,k},f\right)\nonumber \\
 & =\frac{c^{2}}{\left(4\pi d_{g,k}f\right)^{2}}\exp\left[-\mu\left(f\right)d_{g,k}\right],~g=1,2,\label{eq:el-1}
\end{align}
where $\xi_{1,k}^{L}$ and $\xi_{2,k}^{L}$ are the corresponding
THz LOS path-loss, $\mu\left(f\right)$ is the absorption coefficient
at frequency $f$, $d_{1,k}$ is the distance from the BS to the RIS,
$d_{2,k}$ is the distance from the RIS to the user, both for user
$k$, and $c$ is the speed of light. Still referring to (\ref{eq:G})
and (\ref{eq:arj}), $\beta_{1,k,i}^{NL}$ and $\beta_{2,k,i}^{NL}$
denote the corresponding THz NLOS complex gains.

In Eq. (\ref{eq:G}) and (\ref{eq:arj}), $\mathbf{a}_{t,k}\left(\psi_{k},\sigma_{k}\right)$,
and $\mathbf{a}_{r,k}\left(\vartheta_{k},\phi_{k}\right)$ are the
antenna array steering vectors at the \emph{k}-th BS subarray and
$k$-th user, respectively:

\begin{equation}
\begin{aligned}\mathbf{a}_{t,k}\left(\psi_{k},\sigma_{k}\right)\\
= & \frac{1}{\sqrt{m_{t}n_{t}}}\left[1,\ldots,e^{j\frac{2\pi r}{\lambda}[x_{1}\cos\psi_{k}\sin\sigma_{k}+y_{1}\sin\psi_{k}\sin\sigma_{k}]}\right.\\
 & \left.\ldots,e^{j\frac{2\pi r}{\lambda}[(m_{t}-1)\cos\psi_{k}\sin\sigma_{k}+(n_{t}-1)\sin\psi_{k}\sin\sigma_{k}]}\right]^{T},
\end{aligned}
\label{eq:atj}
\end{equation}
where $x_{1}$ and $y_{1}$ denote the index of the BS antenna element,
$0<x_{1}<m_{t}-1$, $0<y_{1}<n_{t}-1$. Additionally, $r$ is the
distance between the BS antenna elements, and $\lambda$ represents
the wavelength of THz signals. Still referring to (\ref{eq:arj}),
we have

\begin{equation}
\begin{aligned}\mathbf{a}_{r,k}\left(\vartheta_{k},\phi_{k}\right)\\
= & \frac{1}{\sqrt{m_{r}n_{r}}}\left[1,\ldots,e^{j\frac{2\pi\gamma}{\lambda}[x_{2}\cos\vartheta_{k}\sin\phi_{k}+y_{2}\sin\vartheta_{k}\sin\phi_{k}]}\right.\\
 & \left.\ldots,e^{j\frac{2\pi\gamma}{\lambda}[(m_{r}-1)\cos\vartheta_{k}\sin\phi_{k}+(n_{r}-1)\sin\vartheta_{k}\sin\phi_{k}]}\right]^{T},
\end{aligned}
\label{eq:arj-13}
\end{equation}
where $x_{2}$ and $y_{2}$ denote the index of the user antenna element,
$0<x_{2}<m_{r}-1$, $0<y_{2}<n_{r}-1$; and $\gamma$ is the distance
between the user antenna elements.

Explicitly, $\mathbf{a}_{sa,k}(\delta_{k},\kappa_{k})$ and $\mathbf{a}_{sd,k}(\varsigma_{k},\varphi_{k})$
in Eq. (\ref{eq:G}) and (\ref{eq:arj}) are the arrival and departure
steering vectors at the $k$-th sub-RIS, respectively. They can be
expressed as follows:

\begin{equation}
\begin{aligned}\mathbf{a}_{sa,k}(\delta_{k},\kappa_{k})\\
= & \frac{1}{\sqrt{m_{s}n_{s}}}\left[1,\ldots,e^{j\frac{2\pi\tau}{\lambda}[x_{s}\cos\delta_{k}\sin\kappa_{k}+y_{s}\sin\delta_{k}\sin\kappa_{k}]}\right.\\
 & \left.\ldots,e^{j\frac{2\pi\tau}{\lambda}[(m_{s}-1)\cos\delta_{k}\sin\kappa_{k}+(n_{s}-1)\sin\delta_{k}\sin\kappa_{k}]}\right]^{T},
\end{aligned}
\label{eq:asaj}
\end{equation}

\begin{equation}
\begin{aligned}\mathbf{a}_{sd,k}(\varsigma_{k},\varphi_{k})\\
= & \frac{1}{\sqrt{m_{s}n_{s}}}\left[1,\ldots,e^{j\frac{2\pi\tau}{\lambda}[x_{s}\cos\varsigma_{k}\sin\varphi_{k}+y_{s}\sin\varsigma_{k}\sin\varphi_{k}]}\right.\\
 & \left.\ldots,e^{j\frac{2\pi\tau}{\lambda}[(m_{s}-1)\cos\varsigma_{k}\sin\varphi_{k}+(n_{s}-1)\sin\varsigma_{k}\sin\varphi_{k}]}\right]^{T},
\end{aligned}
\label{eq:asdj}
\end{equation}
where $x_{s}$ and $y_{s}$ denote the index of the RIS element, $0<x_{s}<m_{s}-1$,
$0<y_{s}<n_{s}-1$; and $\tau$ is the distance between the RIS elements
within the $k$-th sub-RIS.

Therefore, upon combining (\ref{eq:direct}) and (\ref{eq:via ris}),
the signal received by user $k$ from the BS-user and from the BS-RIS-user
channels can be expressed as

\begin{equation}
y_{k}=\bar{y}_{k}+\tilde{y}_{k}.\label{eq:rec sum}
\end{equation}

The THz channels are sparse and only few paths exists \cite{lin_adaptive_2015}.
Moreover, the power difference of the THz signals between the LOS
and NLOS path is significant. Specifically, the power of the first-order
reflected path is attenuated by more than 10 dB on average compared
to the LOS path and that of the second-order reflection by more than
20 dB \cite{lin_adaptive_2015}, so THz channels are LOS-dominant
and the small-scale fading can be ignored. Thus, we will focus on
the LOS path of the THz signal when exploring the channel condition,
while relying on a RIS, and on beamforming schemes. Furthermore, considering
the beamforming gain of the transceivers and the gains of the RIS,
the received power of the BS-user link in (\ref{eq:rec sum}) is much
lower than that of the BS-RIS-user link.

\section{Conditions of Spatial Multiplexing for RIS-aided THz Channels}

In this section, we discuss the conditions of achieving high multiplexing
gains for RISs-aided THz channels. As the propagation of signals at
THz frequencies is ``quasi-optical'', the LOS path dominates the
channel complemented only by a few non-LOS (NLOS) reflected rays due
to the associated high reflection loss. Thanks to the beamforming
gain and flexible placement of RISs, a LOS path may be present between
each pair of the BS subarrays, the sub-RIS and the user's receiver
subarrays. To achieve high multiplexing gains in strong LoS environments
at high frequencies \cite{sarieddeen_terahertz-band_2019}, the antenna
spacing of both the TX and RX together with that of the RIS element
spacing should be set much larger than the operating THz wavelength. 

We first consider the BS-RIS channel. We denote the distances between
two adjacent subarrays at the BS transmitter and two adjacent sub-RIS
respectively by $\chi$ and $\alpha$. Without loss of generality,
we assume symmetry in the remainder of the paper, i.e., $M_{t}=N_{t}=M$,
$M_{s}=N_{s}=N$. Thus, the BS contains $M\times M$ subarrays and
the RIS contains $N\times N$ sub-RISs. The capacity is maximized
when all columns of $\mathbf{G}$ are orthogonal. Hence we formulate
Theorem 1 as follows.
\begin{thm}
\emph{To achieve spatial multiplexing gain for the BS to the RIS line,
the optimal spacing of sub-RISs is}
\begin{equation}
\alpha_{op}=\sqrt{q\frac{d_{1}\lambda}{N}}\label{eq:opalfa}
\end{equation}
\emph{for integer values of $q$, where $d_{1}$ is the distance from
the BS to the RIS, while $\lambda$ is the wavelength of the THz signal.}
\end{thm}
\begin{IEEEproof}
The proof is given in Appendix A.
\end{IEEEproof}
\begin{thm}
\emph{To achieve a high spatial multiplexing gain for the RIS to BS
uplink, the optimal spacing of the BS subarrays is}
\begin{equation}
\chi_{op}=\sqrt{q\frac{d_{1}\lambda}{M}}\label{eq:optx}
\end{equation}
\emph{for integer values of $q$, where $d_{1}$ is the distance from
the RIS to the BS, and $\lambda$ is the wavelength of the THz signal. }
\end{thm}
\begin{IEEEproof}
The proof is similar to that of Theorem 1, hence it is omitted here. 
\end{IEEEproof}
Therefore, condition (\ref{eq:opalfa}) and condition (\ref{eq:optx})
guarantee having an optimal sub-RIS and BS subarray design, respectively. 

The separation of elements in the BS subarrays or sub-RISs may be
achieved via spatial interleaving. Moreover, the required spacing
can be realized by choosing the right elements belonging to each BS
subarray or sub-RIS. Each BS subarray associated with the corresponding
sub-RIS may focus on a specific individual subband of the THz signal
and can also be tuned flexibly according to the different user distances. 

Let us now consider the channels spanning from the sub-RISs to the
users. At the receiver side, each user has a subarray and is generally
separated, so the distance of two adjacent user subarrays is usually
large. Naturally, we cannot impose any constraints on the user positions,
which tend to be random. So we must resort to exploiting the frequency
selectivity of the THz channels for spatial multiplexing. According
to \cite{lin_adaptive_2015}, the THz channels have limited angular
spread of about $40\lyxmathsym{\textdegree}$. Since the beam steering
vectors associated with completely different angles of large-scale
antennas are nearly orthogonal, for the channels spanning from the
\emph{i}th and the \emph{k}th sub-RIS to user\emph{ k, }we have

\begin{equation}
\mathbf{a}_{sd,i}^{H}(\varsigma_{i},\varphi_{i})\mathbf{a}_{sd,k}(\varsigma_{k},\varphi_{k})\simeq0,\quad\varsigma_{i}\neq\varsigma_{k},\varphi_{i}\neq\varphi_{k}.\label{eq:angle}
\end{equation}

As the RIS may be regarded as a large-scale antenna array, the channel
from the RIS to the user can be nearly orthogonal, hence beneficial
spatial multiplexing gains can be achieved. For users that are close
enough to be within the angular spread, the pre-scanning and grouping
technique of \cite{lin_adaptive_2015} can be adopted.

\section{Path-loss of the RIS-aided Near-Field and Far-Field THz Channel}

Based on the assumption that the spherical wave generated by the transmitter
can be approximately regarded as a plane wave at the RIS side when
a transmitter is far away from the RIS, the far-field and near-field
boundary of the antenna array is defined as $D=\frac{2L^{2}}{\lambda}$
, where $D$, $L$ and $\lambda$ represent the distance between the
transmitter and the center of the antenna array, the maximum dimension
of the antenna array and the wavelength of the signal, respectively
\cite{tang_wireless_2021}.\emph{ }For the sub-RIS case, we have

\begin{equation}
D=\frac{2N^{2}m_{s}n_{s}d_{x}d_{y}}{\lambda}.
\end{equation}

Generally, $d_{x}=d_{y}=\frac{\lambda}{2}$, so we get

\begin{equation}
D=\frac{N^{2}m_{s}n_{s}\lambda}{2}.\label{eq:D}
\end{equation}

Let us now discuss the THz channel path-loss in the near-field and
far-field of a RIS, respectively. Let us represent the RIS-aided cascaded
channel by

\begin{equation}
\mathbf{T}_{k}=\mathbf{H}_{k}\mathbf{O}\mathbf{G}.
\end{equation}

\begin{thm}
For near-field beamforming over the RIS-aided THz channel, the path-loss
of $\mathbf{T}_{k}$ can be expressed by

\begin{equation}
\zeta_{k}=\frac{c^{2}}{\left(4\pi f\right)^{2}\left(d_{1}+d_{2}\right)^{2}}e^{-\mu(f)\left(d_{1}+d_{2}\right)}.\label{eq:near}
\end{equation}
\end{thm}
\begin{IEEEproof}
Based on the path-loss of the near-field RIS-aided beamforming \cite{tang_wireless_2021}
and the THz channel (\ref{eq:el-1}), the path-loss of the cascaded
BS-RIS-user channel $\mathbf{T}_{k}$ can be formulated as
\begin{align*}
\zeta_{k} & =\xi^{L}\left(d_{1}+d_{2},f\right)\\
 & =\frac{c^{2}}{\left(4\pi f\right)^{2}\left(d_{1}+d_{2}\right)^{2}}e^{-\mu(f)\left(d_{1}+d_{2}\right)}.
\end{align*}
Explicitly, the signal transmission process is equivalent to that
of a signal traveling the distance $d_{1}+d_{2}$.
\end{IEEEproof}
\begin{thm}
For far-field beamforming over the RIS-aided THz channel, the path-loss
of $\mathbf{T}_{k}$ can be expressed as

\begin{equation}
\hat{\zeta}_{k}=\frac{c^{2}}{\left(4\pi f\right)^{2}d_{1}^{2}d_{2}^{2}}e^{-\mu(f)\left(d_{1}+d_{2}\right)}.\label{eq:far}
\end{equation}
\end{thm}
\begin{IEEEproof}
According to the path-loss of the far-field RIS-aided beamforming
\cite{tang_wireless_2021} and the THz channel (\ref{eq:el-1}), the
path-loss of the cascaded BS-RIS-user channel $\mathbf{T}_{k}$ can
be formulated as
\[
\hat{\zeta}_{k}=\frac{c^{2}}{\left(4\pi f\right)^{2}d_{1}^{2}d_{2}^{2}}e^{-\mu(f)\left(d_{1}+d_{2}\right)}.
\]
\end{IEEEproof}

\section{Sensor-based Channel Estimation Scheme}

In this section, we propose a location-aided channel estimation scheme.
The location information obtained by the sensor allows us to expedite
the channel estimation and beamforming processes. This is the first
location-aware channel estimator for THz channels. 

Given the RIS-aided THz channel models of Section II, estimating the
RIS-aided channel is equivalent to inferring the parameters of the
channel paths; namely the AoA, the AoD, and the path-loss of each
path. The RIS locations are generally known by the BS. Four UWB sensors
are integrated into the four corners of the RIS, as shown in Fig.
\ref{fig:System-model-of}, which is the minimum number of UWB nodes
required for supporting 3D localization. Thanks to the ultra-low power
density and 3.1-10.6 GHz frequency band of UWB, the interference between
the UWB ranging signal and the THz signals is negligible. The distance
between each UWB sensor and the user can be estimated using either
a \emph{Time-of-Arrival} (TOA) or \emph{Two-Way Time of-Flight }(TW-ToF)
based ranging method \cite{lazzari_numerical_2017}. However, they
introduce the ranging error $\varepsilon$. Then, the 3D user position
can be determined by multi-lateration algorithm upon solving a set
of nonlinear equations:

\begin{equation}
\left[x_{s_{i}}-x_{U}\right]^{2}+\left[y_{s_{i}}-y_{U}\right]^{2}+\left[z_{s_{i}}-z_{U}\right]^{2}=\tilde{d}_{i}^{2}\label{eq:UWB}
\end{equation}
where $i=\{1,2,3,4\}$, $P_{s_{i}}=\left[x_{s_{i}},y_{s_{i}},z_{s_{i}}\right]$
indicates the coordinates of each UWB node on the RIS, $P_{U}=\left[x_{U},y_{U},z_{U}\right]$
is the unknown user position, and $\tilde{d}_{i}=d_{i}+\varepsilon=\left|P_{U}-P_{s_{i}}\right|+\varepsilon$
is the measurement of the distance between the UWB sensors and the
UE. Given the ranging error $\varepsilon$, an approximate of (\ref{eq:UWB})
must be used instead of the intersection of four spheres at a single
point found in the ideal scenario \cite{lazzari_numerical_2017}. 

We then adopt the user location information provided by the UWB sensors
to obtain angle and path-loss of the LOS path. The LOS path-loss can
be readily calculated from the BS-user and from the RIS-user distance
according to (\ref{eq:near}) and (\ref{eq:far}). The AOD/AOA of
the LOS path can be inferred by trigonometry as follows.

Without loss of generality, the $k$-th BS subarray and the $k$-th
sub-RIS of Fig. \ref{fig:System-model-of} are assumed to be located
at $(x_{B,k},y_{B,k},z_{B,k})$ and $(0,0,0)$, respectively. Then
the effective AOA at the $k$-th sub-RIS from the $k$-th BS subarray
can be calculated as

\begin{equation}
\delta_{k}=\arctan\left(\frac{y_{B,k}}{x_{B,k}}\right),
\end{equation}

\begin{equation}
\kappa_{k}=\arcsin\left(\frac{z_{B,k}}{d_{1,k}}\right),
\end{equation}
where $d_{1,k}=\sqrt{x_{B,k}^{2}+y_{B,k}^{2}+z_{B,k}^{2}}$. 

Similarly, the effective AOD from the $k$-th BS subarray to the $k$-th
sub-RIS are:

\begin{equation}
\psi_{k}=-\delta_{k},
\end{equation}

\begin{equation}
\sigma_{k}=-\kappa_{k}.
\end{equation}

Therefore, according to (\ref{eq:atj}) and (\ref{eq:asaj}), $\mathbf{a}_{t,k}\left(\psi_{k},\sigma_{k}\right)$
and $\mathbf{a}_{sa,k}(\delta_{k},\kappa_{k})$ can be obtained.

Let $\left(\hat{x}_{U,k},\hat{y}_{U,k},\hat{z}_{U,k}\right)$ denote
the estimated location of user \emph{k} obtained by the UWB sensors.The\emph{
k}-th sub-RIS calculates its effective AOD from itself to the\emph{
k}-th user as

\begin{equation}
\hat{\varsigma}_{k}=\arctan\left(\frac{\hat{y}_{\mathrm{\mathit{U}},k}}{\hat{x}_{U,k}}\right),\label{eq:RIS1}
\end{equation}

\begin{equation}
\hat{\varphi}_{k}=\arcsin\left(\frac{\hat{z}_{\mathrm{\mathit{U}},k}}{\hat{d}_{2,k}}\right),\label{eq:RIS2}
\end{equation}
where $\hat{d}_{2,k}=\sqrt{\hat{x}_{U,k}^{2}+\hat{y}_{U,k}^{2}+\hat{z}_{U,k}^{2}}$.

Similarly, the\emph{ k}-th user calculates its effective AOA from
itself to the\emph{ k}-th sub-RIS as

\begin{equation}
\hat{\vartheta}_{k}=-\hat{\varsigma}_{k},\label{eq:UE1}
\end{equation}

\begin{equation}
\hat{\phi}_{k}=-\hat{\varphi}_{k}.\label{eq:UE2}
\end{equation}

Thus, according to (\ref{eq:asdj}) and (\ref{eq:arj-13}), $\mathbf{a}_{sd,k}(\hat{\varsigma}_{k},\hat{\varphi}_{k})$
and $\mathbf{a}_{r,k}\left(\hat{\vartheta}_{k},\hat{\phi}_{k}\right)$
can be obtained.

Furthermore, by substituting $d_{1,k}$ and $\hat{d}_{2,k}$ into
(\ref{eq:el-1}), we can obtain the channel gain $\beta_{1,k}^{L}$
and $\hat{\beta}_{2,k}^{L}$.

\subsection{Analog Active Beamforming Relying on User Locations}

Then, the estimated angle information can be used to design the BS
subarray's active beamforming and sub-RIS's passive beamforming. However,
because the active and the passive beamforming are coupled, the optimization
problem is nonconvex, so the global optimum is typically a challenge
to find. Although some suboptimal algorithms have been proposed for
MISO systems relying on alternating optimization \cite{guo_weighted_2020},
the computational complexity of these algorithms is excessive because
of the large number of reflection elements. Hence we circumvent this
challenge by opting for the low-complexity separate optimization of
the BS subarray and of each sub-RIS, because then convenient closed-form
solutions can be obtained. Specifically, the BS utilizes the angle
information of the BS-RIS link to design the active beamforming. Without
loss of generality, we assume that the \emph{k}-th user is assisted
by the \emph{k}-th sub-RIS. Thus, the active beamforming designed
for the $k$-th user should be aligned to the $k$-th sub-RIS. As
such, the transmit beam of the $k$-th BS subarray is designed as 

\emph{
\begin{equation}
\mathbf{w}_{k}=\sqrt{\frac{p_{k}}{m_{t}n_{t}}}\mathbf{a}_{t,k}\left(\psi_{k},\sigma_{k}\right),\label{eq:wj}
\end{equation}
}where $p_{k}$ is the transmit power of the \emph{k}-th BS subarray.
Therefore, the channel between the BS and RIS can be readily estimated
with the aid of the previously calibrated accurate angles, and we
would only focus our attention on the estimation of the time-variant
RIS-UE channels.

The received beam of the $k$-th user is given by
\begin{equation}
\mathbf{\hat{v}}_{k}=\sqrt{\frac{1}{m_{r}n_{r}}}\mathbf{a}_{r,k}\left(\hat{\vartheta}_{k},\hat{\phi}_{k}\right).\label{eq:v}
\end{equation}

\subsection{RIS Phase Shift Based Passive Beamforming Relying on User Locations}

We then use the estimated angle information of the RIS-user link for
the RIS phase shift based passive beamforming design. The $k$-th
user's received signal is maximized with the \emph{k}-th sub-RIS by
optimizing the phase shift beam $\mathbf{O}_{k}$. The optimization
problem can be formulated as:

\emph{
\begin{equation}
\begin{array}{c}
\max_{\mathbf{O}_{k}}\left|\mathbf{v}_{k}^{H}\mathbf{H}_{k}\mathbf{O}\mathbf{G}\mathbf{Wf_{\mathit{k}}}\right|^{2},\\
\text{ s.t. }\left|\left[\mathbf{O}_{k}\right]_{i,i}\right|=1,\quad i=1,\ldots,m_{s}n_{s}.
\end{array}\label{eq:max}
\end{equation}
}According to $\mathbf{y}^{H}\mathbf{X}\mathbf{z}=\mathbf{x}^{T}(\mathbf{y^{*}}\odot\mathbf{z})$
with $\mathbf{X}=\mathrm{diag}(\mathbf{x})$, we have

\begin{equation}
\begin{aligned}\mathbf{v}_{k}^{H}\mathbf{H}_{k}\mathbf{O}\mathbf{G}\mathbf{W}\mathbf{\mathbf{f_{\mathit{k}}}} & =\mathbf{q}_{k}^{T}\left[\left(\mathbf{H}_{k}^{T}\mathbf{v}_{k}^{*}\right)\odot\mathbf{G}\mathbf{W}\mathbf{f_{\mathit{k}}}\right]\\
 & =\mathbf{q}_{k}^{T}\left[\left(\mathbf{H}_{k}^{T}\mathbf{v}_{k}^{*}\right)\odot\mathrm{\mathbf{a}}_{sa,k}\right]\mathbf{\mathrm{\mathbf{a}}}_{t,k}^{H}\mathbf{W}\mathbf{f_{\mathit{k}}}.
\end{aligned}
\end{equation}
Thus, the objective function in (\ref{eq:max}) becomes

\begin{equation}
\left|\mathbf{q}_{k}^{T}\left[\left(\mathbf{H}_{k}^{T}\mathbf{v}_{k}^{*}\right)\odot\mathrm{\mathbf{a}}_{sa,k}\right]\right|^{2}\left|\mathbf{\mathrm{\mathbf{a}}}_{t,k}^{H}\mathbf{W}\mathbf{f_{\mathit{k}}}\right|^{2}.
\end{equation}
According to (\ref{eq:wfj}) and (\ref{eq:wj}), $\left|\mathbf{\mathrm{\mathbf{a}}}_{t,k}^{H}\mathbf{W}\mathbf{f_{\mathit{k}}}\right|^{2}$
is a constant independent of $\mathbf{q}_{k}$, so the optimization
problem equals to 

\begin{equation}
\begin{array}{c}
\max_{\mathbf{q}_{k}^{T}}\left|\mathbf{q}_{k}^{T}\left[\left(\mathbf{H}_{k}^{T}\mathbf{v}_{k}^{*}\right)\odot\mathrm{\mathbf{a}}_{sa,k}\right]\right|^{2},\\
\text{ s.t. }\left|\left[\mathbf{q}_{k}\right]_{i}\right|=1,\quad i=1,\ldots,m_{s}n_{s}.
\end{array}
\end{equation}
Hence, the solution of the above optimization problem is

\begin{equation}
\mathbf{q}_{k}=\left[\left(\mathbf{H}_{k}^{T}\mathbf{v}_{k}^{*}\right)\odot\mathrm{\mathbf{a}}_{sa,k}\right]^{*}.
\end{equation}
Using the estimated angles calculated from the position information
obtained by the UWB sensors, we design the RIS phase shift beam as 

\begin{equation}
\mathbf{\hat{q}}_{k}=\left[\left(\hat{\mathbf{\bar{H}}}_{k}^{T}\mathbf{\hat{v}}_{k}^{*}\right)\odot\boldsymbol{\mathrm{a}}_{sa,k}\right]^{*}.\label{eq:w}
\end{equation}
where $\hat{\bar{\mathbf{H}}}_{k}=\sqrt{m_{r}n_{r}m_{s}n_{s}}\hat{\beta}_{2,k}^{L}\mathbf{a}_{r,k}\left(\hat{\vartheta}_{k},\hat{\phi}_{k}\right)\mathbf{a}_{sd,k}^{H}\left(\hat{\varsigma}_{k},\hat{\varphi}_{k}\right).$

Based on (\ref{eq:v}) and (\ref{eq:w}), both the phase shift based
beamforming and the UE's receiver beamforming both depend on the angles
calculated from the estimated user location information obtained by
the UWB sensors, which contains positioning errors. The inevitable
positioning errors will result in passive transmit beamforming misalignment
with the user, which is further aggravated by the UE's receiver beamforming
misalgnment with the RIS' transmit beam. Given the extremely narrow
pencil beam characteristics of THz signals, the system's performance
will thus be severely degraded. 

Without being constrained by the positioning errors of the UWB sensors,
we further propose the \emph{Precise Beamforming Algorithm} (PBA)
for the joint RIS phase shift based beamforming and for the UE's receiver
beamforming design.

\subsection{Precise Beamforming Algorithm for joint RIS Phase Shift and UE Receiver
Beamforming}

\begin{figure}[tbh]
\begin{centering}
\includegraphics[scale=0.5]{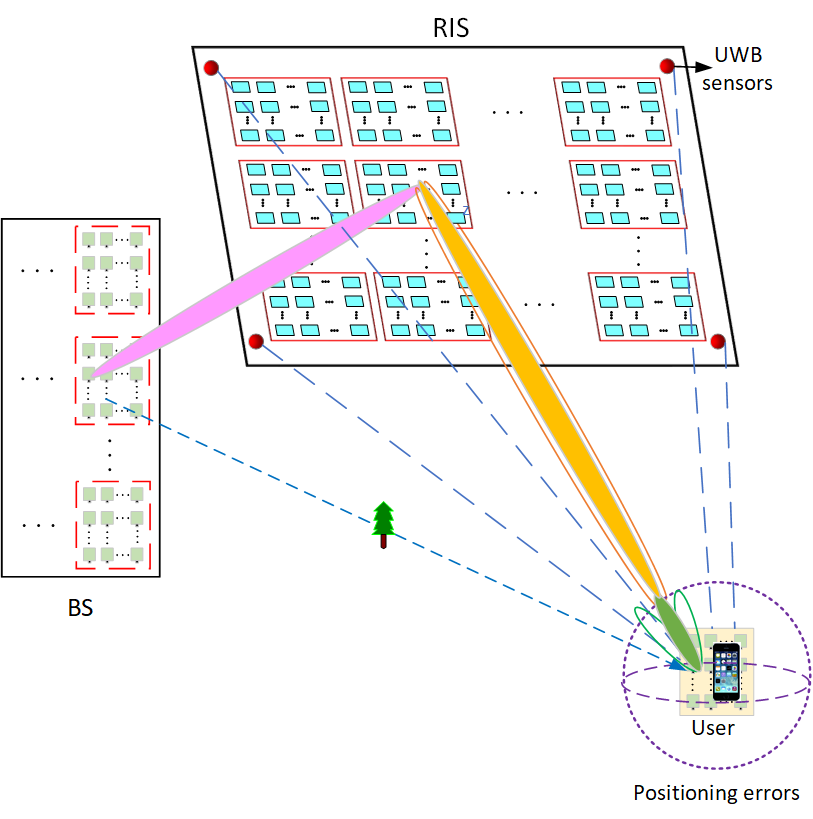}
\par\end{centering}
\caption{\label{fig:Precise-Beamforming-(PBF)}Precise Beamforming (PBF) for
the RIS phase shift beamforming design (RIS-BF) and UE receiver beamforming
(RBF) design based on the user location.}
\end{figure}

As shown in Fig. \ref{fig:Precise-Beamforming-(PBF)}, we propose
the PBA for the improved alignment of the RIS's phase shift based
beamforming and the UE's receiver beamforming. The accurate position
of the $k$th UE denoted by $\left(x_{U,k},y_{U,k},z_{U,k}\right)$
can be assumed to be uniformly distributed within a sphere with the
radius $r_{e}$ (for UWB positioning, generally $r_{e}\leq0.1$ m)
and center $\left(\hat{x}_{U,k},\hat{y}_{U,k},\hat{z}_{U,k}\right).$

The algorithm is formulated as follows:

\begin{algorithm}[tbh]
\caption{Precise Beamforming Algorithm (PBA) for the RIS phase shift beamforming
design and UE receiver beamforming design }

1: \textbf{Input: }The receiver beamforming (RBF) codebook $\mathscr{V}$
at the UE and the phase shift based beamforming codebook $\mathscr{W}$
at the RIS (RIS-BF). 

2: Locate all the users by the UWB sensors.

3: \textbf{For} $k=1:K$ \textbf{do} 

4: According to (\ref{eq:w}), obtain the initial RIS-BF for the $k$th
sub-RIS.

5: According to (\ref{eq:v}), obtain the initial RBF for user $k$. 

6: Select the sub-codebook from $\mathscr{V}$ to obtain $\mathscr{\widetilde{V}}_{k}$
within the range of the UWB positioning errors.

7: Select the sub-codebook from $\mathscr{\mathscr{W}}$ to obtain
$\mathscr{\widetilde{\mathscr{W}}}_{k}$ within the range of the UWB
positioning errors.

\textbf{~~~Repeat }

8: Search $\mathscr{\widetilde{V}}_{k}$ to find the optimal RBF $\mathbf{v}_{k}$
for the $k$th UE so that $\left\{ \mathbf{a}_{r,k}\left(\vartheta_{k},\phi_{k}\right)\right\} =\underset{\mathbf{v}_{k}}{argmax}\left|\mathbf{v}_{k}^{H}\mathbf{H}_{k}\right|^{2}$.

9: Search $\mathscr{\widetilde{\mathscr{W}}}_{k}$ to find the optimal
RIS-BF $\mathbf{O}_{k}$ for the $k$th sub-RIS so that $\left\{ \mathbf{a}_{sd,k}(\varsigma_{k},\varphi_{k})\right\} =\underset{\mathbf{O}_{k}}{argmax}\left|\mathbf{v}_{k}^{H}\mathbf{H}_{k}\mathbf{O}_{k}\right|^{2}$.

\textbf{~~~until $\left|\mathbf{v}_{k}^{H}\mathbf{H}_{k}\mathbf{O}_{k}\right|^{2}$
}is maximized.

10: \textbf{end for}

11:\textbf{ Output:} $\mathbf{O}_{k}$ and $\mathbf{v_{\mathit{k}}}$,
$k=1,2,...,K$.
\end{algorithm}

It is worth mentioning that the results of $\mathbf{O}_{k}$ may also
be fed back to the UWB positioning system for improving its accuracy,
which is also a further benefit of joint sensing and communication.

\subsection{Complexity Analysis}

In this subsection, we compare the search complexity of our proposed
schemes to those of other beam training schemes \cite{ning_terahertz_2021}.
The results are shown in Table II. Compared to other schemes, the
search time of our proposed scheme is negligible and unrelated to
$N$, thanks to the high-precision UWB positioning. By contrast, the
complexity of the benchmarks increases. Therefore, the proposed scheme
is eminently suitable for time-varying user positions or delay-sensitive
applications.

\begin{table*}[tbh]
\caption{Comparison of beam training schemes}

\centering{}%
\begin{tabular}{|>{\centering}p{5cm}|>{\centering}p{3cm}|>{\centering}p{3cm}|>{\centering}p{5cm}|}
\hline 
\raggedright{}Beam training schemes &
Applicable to RIS-aided system &
Applicable to THz &
Search time for RIS-aided system\tabularnewline
\hline 
\hline 
\raggedright{}Exhaustive search &
Yes &
Yes &
$N^{2}+N^{4}$\tabularnewline
\hline 
\raggedright{}One-side search \cite{nitsche_ieee_2014} &
No &
No &
$-$\tabularnewline
\hline 
\raggedright{}Adaptive binary-tree search \cite{alkhateeb_channel_2014} &
No &
No &
$-$\tabularnewline
\hline 
\raggedright{}Two-stage training scheme \cite{lin_subarray-based_2017} &
No &
Yes &
$-$\tabularnewline
\hline 
\raggedright{}Tree dictionary (TD) and PS deactivation (PSD) codebook
based search \cite{ning_terahertz_2021} &
Yes &
Yes &
$18N+12\log_{3}N-3$ (TD) or $6N+4\log_{3}N-1$ (PSD)\tabularnewline
\hline 
\raggedright{}Proposed sensor-based scheme with PBA &
Yes &
Yes &
Negligible and not related to $N$\tabularnewline
\hline 
\end{tabular}
\end{table*}

\subsection{Design of the BS's Digital Precoder }

Following the beam training and channel estimation using sensor based
PBA, we will now design the digital \emph{transmit precoder} (TPC)
for interference cancellation among different users. Given the effective
channel $\hat{\mathbf{G}}$ estimated above, the digital TPC can be
designed as follows. Let

\begin{equation}
\mathbf{\hat{T}}_{k}=\mathbf{\hat{v}}_{k}^{H}\hat{\bar{\mathbf{H}}}_{k}\mathbf{\hat{O}}\mathbf{G}\mathbf{W}.
\end{equation}
Specifically, the \emph{minimum mean squared error} (MMSE) TPC is
formulated as

\begin{equation}
\mathbf{F}=\left[\mathbf{f}_{1},\mathbf{f}_{2},\ldots,\mathbf{f}_{K}\right]=\left[\left(\hat{\mathbf{T}}^{H}\hat{\mathbf{T}}+\frac{K\sigma^{2}}{P_{s}}\mathbf{W}^{H}\mathbf{W}\right)^{-1}\hat{\mathbf{T}}^{H}\right]^{-1}.
\end{equation}
The \emph{Zero-forcing} (ZF) digital TPC employed as a benchmark is
formulated as: 

\begin{equation}
\mathbf{F}=\left[\mathbf{f}_{1},\mathbf{f}_{2},\ldots,\mathbf{f}_{K}\right]=\hat{\mathbf{T}}^{H}\left(\hat{\mathbf{T}}\hat{\mathbf{T}}^{H}\right)^{-1}\Delta=\hat{\mathbf{F}}\Delta,
\end{equation}
where $\hat{\mathbf{F}}=\left[\hat{\mathbf{f}}_{1},\mathbf{\hat{f}}_{2},\ldots,\mathbf{\hat{f}}_{K}\right]\triangleq\hat{\mathbf{T}}^{H}\left(\hat{\mathbf{T}}\hat{\mathbf{T}}^{H}\right)^{-1}$,
and $\Delta$ is a diagonal matrix representing the digital TPC power
so that $\left\Vert \mathbf{Wf}_{k}\right\Vert ^{2}=1,k\in\mathcal{K}.$
Particularly, the \emph{k}-th diagonal element of $\Delta$ is given
by $\Delta_{k,k}=\frac{1}{\left\Vert \mathbf{W}\hat{\mathbf{f}}_{k}\right\Vert }$. 

\section{Performance Analysis}

Based on the received signal (\ref{eq:rec sum}), in this section,
we characterize the system performance in terms of the achievable
sum-rate of the users expressed by

\begin{equation}
R=\mathbb{E}\left\{ \stackrel[k=1]{K}{\sum}r_{k}\right\} ,
\end{equation}
where we have:

\begin{equation}
r_{k}=\log_{2}\left(1+\frac{\frac{P}{K}\left|\mathbf{v}_{k}^{H}\mathbf{H}_{k}\mathbf{O}\mathbf{G}\mathbf{W}\mathbf{f}_{k}\right|^{2}}{\frac{P}{K}\sum_{i\neq k}^{K}\left|\mathbf{v}_{i}^{H}\mathbf{H}_{i}\mathbf{\mathbf{O}\mathbf{G}W}\mathbf{f}_{i}\right|^{2}+\left|\mathbf{v}_{k}^{H}\mathbf{n}_{k}\right|^{2}}\right).
\end{equation}
According to the orthogonality of different users arranged by the
sub-RIS and subarray of the BS, which is achieved by the conditions
shown in Section III, as well as the previous digital TPC, we have

\begin{equation}
\frac{P}{K}\sum_{i\neq k}^{K}\left|\mathbf{v}_{i}^{H}\mathbf{H}_{i}\mathbf{O}\mathbf{G}\mathbf{W}\mathbf{f}_{i}\right|^{2}\approx0.
\end{equation}
Therefore, we arrive at:

\begin{equation}
r_{k}=\log_{2}\left(1+\frac{\frac{P}{K}\left|\mathbf{v}_{k}^{H}\mathbf{H}_{k}\mathbf{O}\mathbf{G}\mathbf{W}\mathbf{f}_{k}\right|^{2}}{\left|\mathbf{v}_{k}^{H}\mathbf{n}_{k}\right|^{2}}\right).
\end{equation}

\section{Simulation Results and Discussions}

In this section, we evaluate the performance of our sensor-based channel
estimation scheme proposed for RIS-aided THz MIMO systems. We consider
both the near-field and far-field of RIS-BF. There are 4 users having
different positions. At the BS, each RF chain is connected to a single
sub-arrray and each subarray corresponds to a user. As there is a
low-attenuation THz transmission window at 350 GHz \cite{moldovan_and_2014},
we consider the THz frequency band at 350 GHz in the simulations.
The number of channels between the user and the BS is set to 3 due
to the sparsity of the THz channels \cite{lin_adaptive_2015}. The
simulation parameters are shown in Table III. The user locations are
generated randomly and the channel conditions of (\ref{eq:angle})
are satisfied. For the system operating without RIS, the \emph{channel
state information} (CSI) is assumed to be perfectly estimated by the
BS. For the RIS-aided system, the BS initially does not know the CSI
and the channel estimation is conducted based on our proposed scheme.
The simulations are conducted in Matlab and 100,000 random runs are
performed.\textbf{ }

\begin{table}[tbh]
\caption{Simulation Parameters}

\begin{tabular}{>{\raggedright}m{25mm}|>{\centering}m{15mm}|>{\raggedleft}m{35mm}}
\hline 
Parameters &
Symbols &
Values\tabularnewline
\hline 
Center frequency of the subband &
$f_{1}$ &
$350$ GHz\tabularnewline
\hline 
Noise power &
$N_{0}$ &
\textminus 75 dBm\tabularnewline
\hline 
Bandwidth &
$W$ &
1 GHz\tabularnewline
\hline 
Number of Users &
$K$ &
4\tabularnewline
\hline 
Location of the center of the RIS &
 &
(0, 0, 0)\tabularnewline
\hline 
Location of the center of the BS in the nearfield case &
 &
(-0.6, -0.7, 0.4)\tabularnewline
\hline 
Locations of the users in the nearfield case &
 &
(0, 0, 1.1), (0, 0.45, 0.22), (0.742, 0, 0.3), (0.71, 0.7, 0.1)\tabularnewline
\hline 
Location of the center of the BS in the farfield case &
 &
(-4, -4, -2)\tabularnewline
\hline 
Locations of the users in the farfield case &
 &
(2, 2, 1), (0, 3.4, 0.85), (4, 0, 2.07), (0, 0, 5.8)\tabularnewline
\hline 
\end{tabular}
\end{table}

\begin{figure}[tp]
\begin{centering}
\textsf{}%
\subfloat[]%
{\textsf{\includegraphics[scale=0.55]{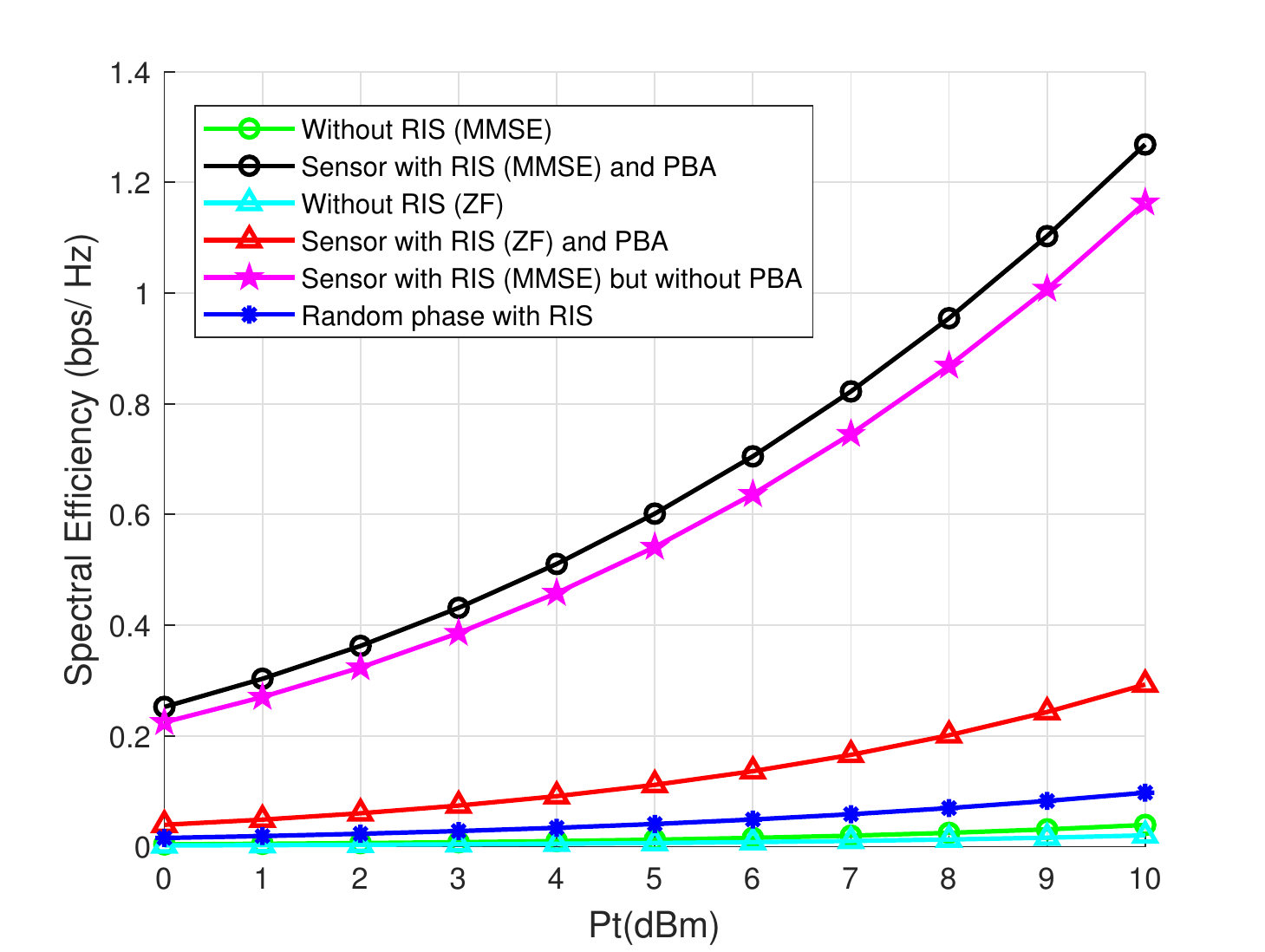}}}
\par\end{centering}
\begin{centering}
\subfloat[]%
{\includegraphics[scale=0.55]{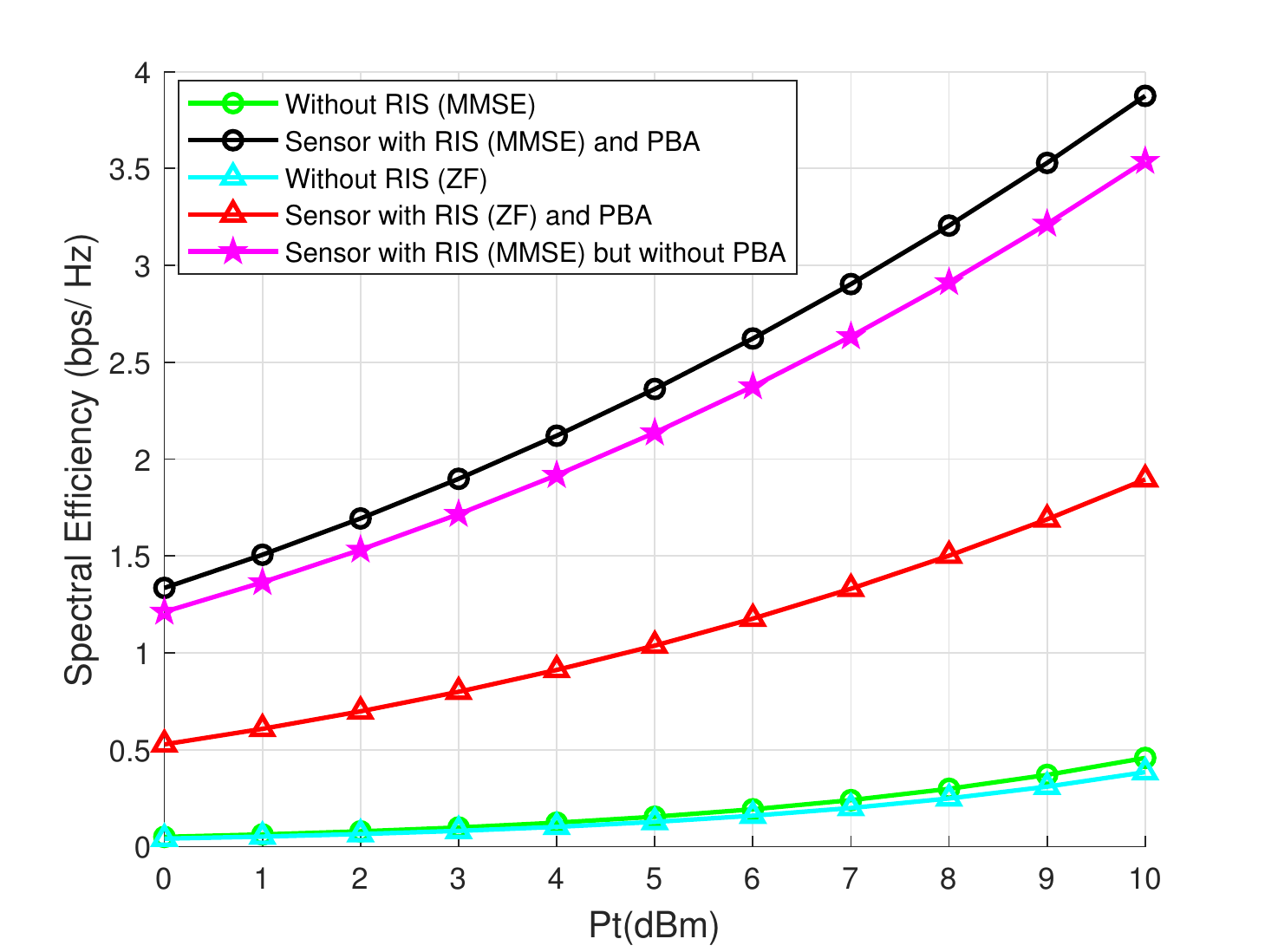}}
\par\end{centering}
\begin{centering}
\subfloat[]%
{\includegraphics[scale=0.55]{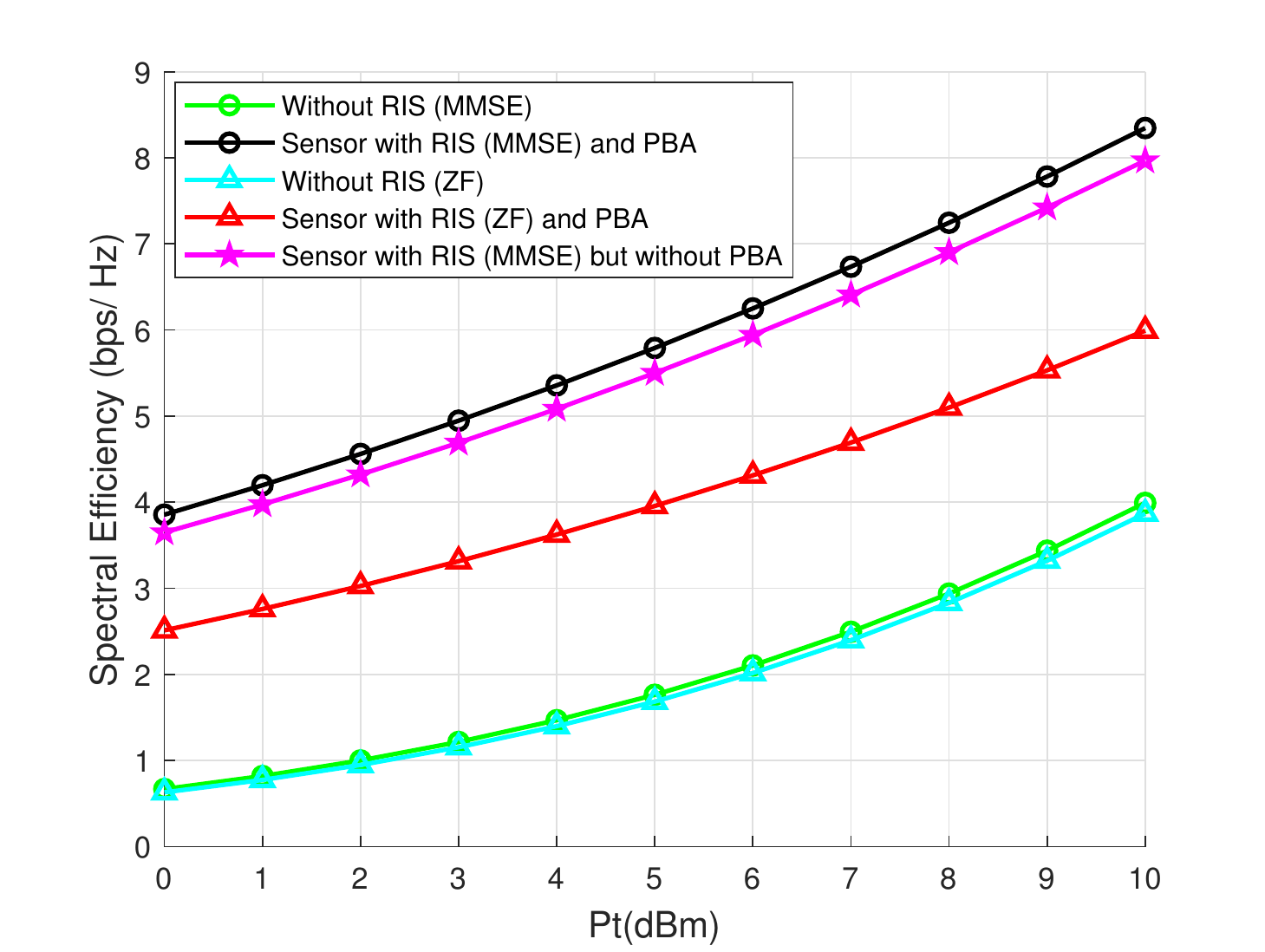}}
\par\end{centering}
\caption{\label{fig:Achievable-rate-(bps/Hz)}Achievable rate (bps/Hz) versus
transmit power (dBm) in Far-field case with 4 BS UPA subarrays, 4
UPA sub-RISs and 4 users: (a) BS subarray $4\times4$, sub-RIS $4\times4$,
user UPA $4\times4$; (b) BS subarray $8\times8$, sub-RIS $8\times8$,
user UPA $8\times8$; (c) BS subarray $16\times16$, sub-RIS $16\times16$,
user UPA $16\times16$.}
\end{figure}

For the far-field scenario, we first assume that each BS antenna subarray
is a $4\times4$ UPA and each user is equipped with one RF chain associated
with a $4\times4$ UPA subarray. Each sub-RIS is also a $4\times4$
UPA. According to (\ref{eq:D}), we have $D\approx0.027$ m. Then
we increase the BS antenna subarray, RIS subarray and user subarray
to $8\times8$ UPA and $16\times16$ UPA. The boundary is $D\approx0.11$
m and $D\approx0.44$ m, correspondingly. Therefore, as both the BS
and users are in the far-field of the RIS, the locations of the center
of the BS arrays and of the RIS are generated randomly, and shown
in Table III. For the RIS-aided channel, the distance between the
BS and the RIS is 6 m while the distances between the RIS and the
users are 3 m, 4.5 m, 3.5 m and 5.8 m, respectively. The simulation
results are shown in Fig. \ref{fig:Achievable-rate-(bps/Hz)} (a)-(c).
Observe that without the proposed sensor based channel estimation
the performance of RIS-aided channel associated with random phase
is only a little better than that without the RIS. For comparison,
our proposed scheme achieves much higher spectral efficiency than
the original system operating without the RIS. Moreover, the sensor
based PBA scheme achieves higher spectral efficiency than that without
it, because the PBA further eliminates the effect of sensor positioning
errors. Additionally, the proposed MMSE scheme outperforms its ZF
counterpart since ZF may cause noise amplification. Thus, we only
characterize the proposed MMSE-based scheme in the following simulations. 

\begin{figure}[tp]
\begin{centering}
\subfloat[]%
{\includegraphics[scale=0.55]{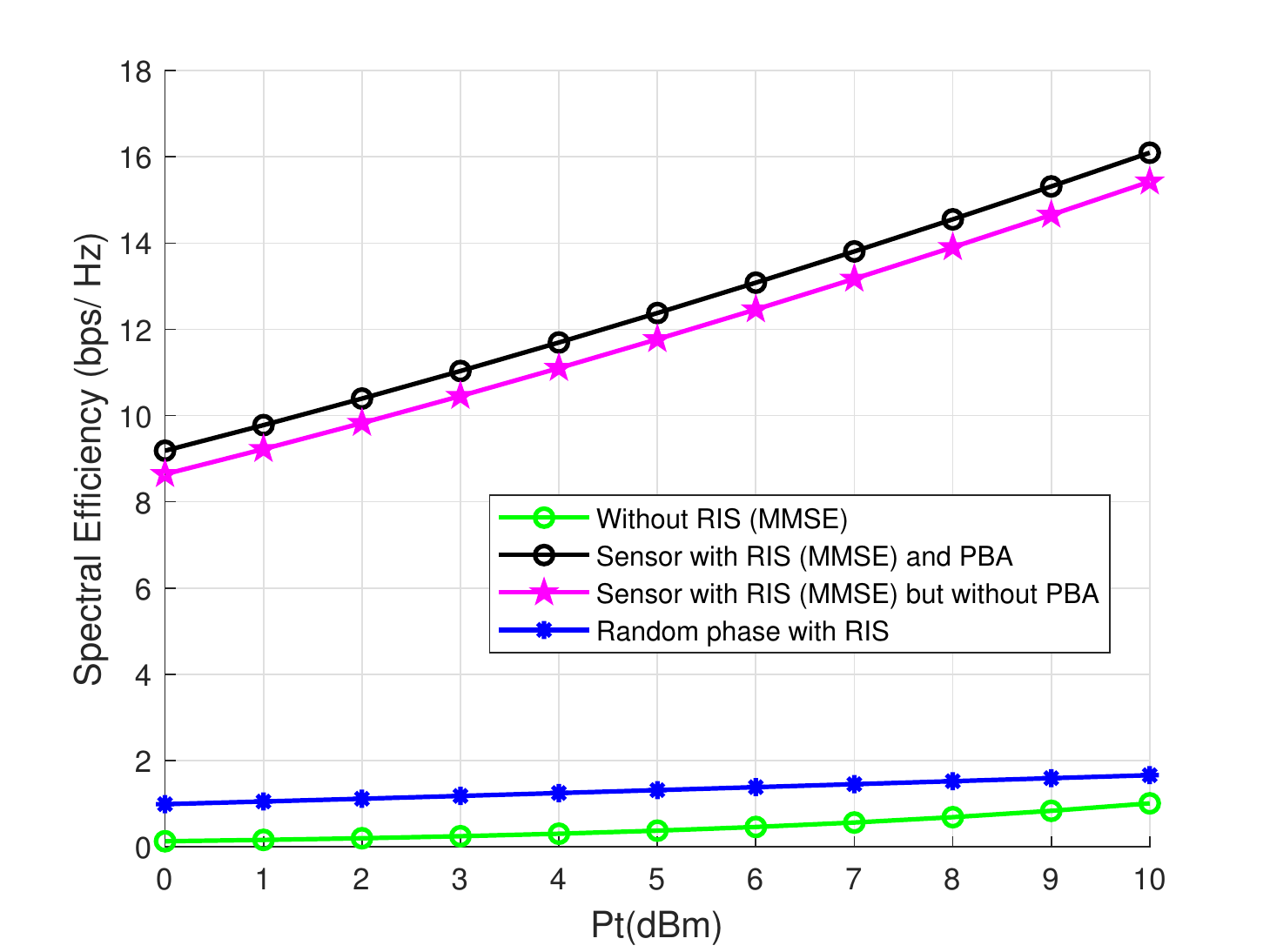}

}
\par\end{centering}
\begin{centering}
\subfloat[]%
{\includegraphics[scale=0.55]{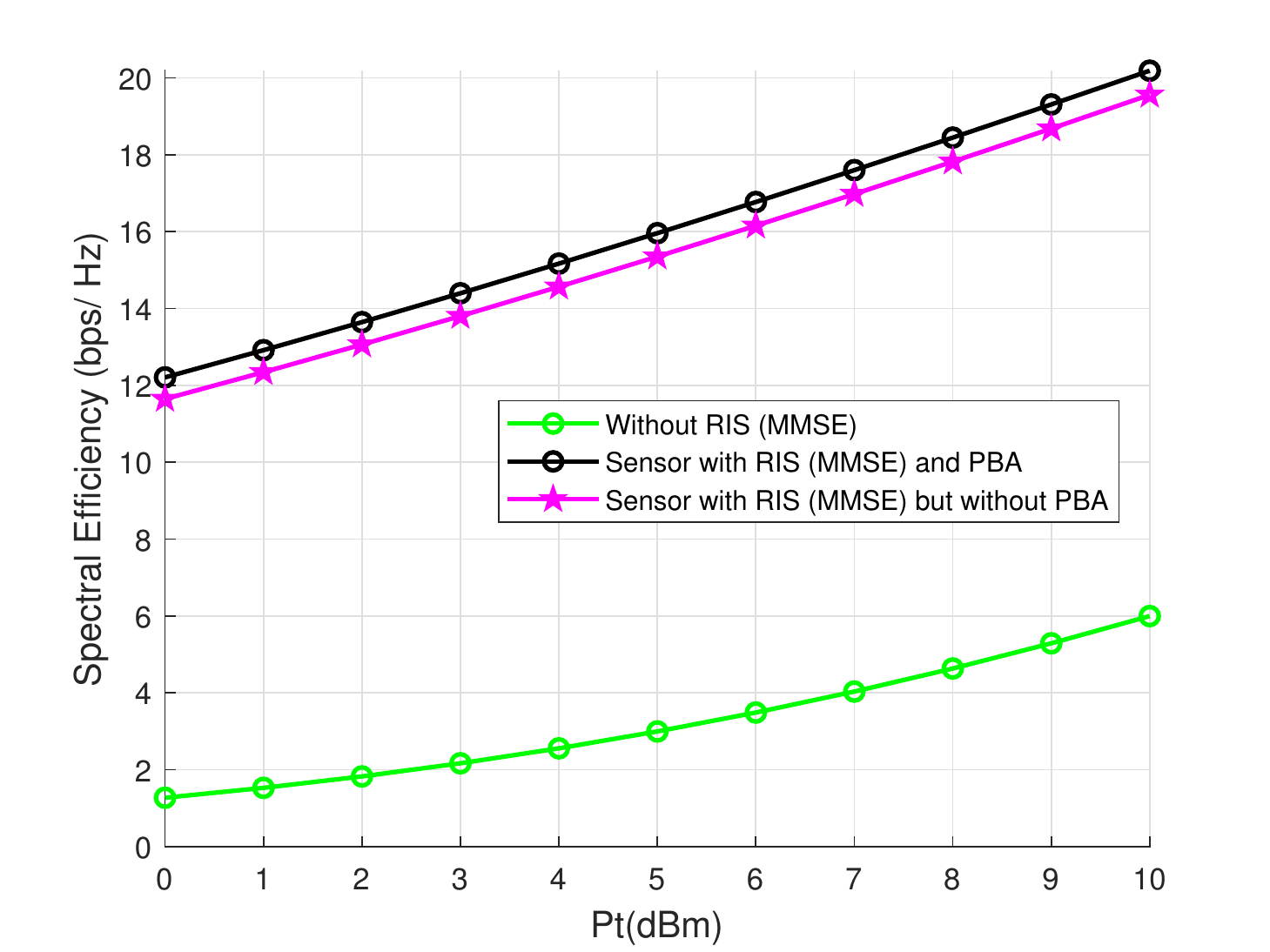}

}
\par\end{centering}
\caption{\label{fig:4 Achievable-rate-(bps/Hz)-1}Achievable rate (bps/Hz)
versus transmit power (dBm) in Near-field case with 4 BS UPA subarrays,
4 UPA sub-RISs and 4 users: (a) BS subarray $4\times4$, sub-RIS $32\times32$,
user UPA $4\times4$; (b) BS subarray $8\times8$, sub-RIS $32\times32$,
user UPA $8\times8$.}
\end{figure}

For the near-field scenario, each sub-RIS is a $32\times32$ UPA.
According to (\ref{eq:D}), the boundary is $D\approx1.76$ m. Then
we increase the dimensions of the BS subarray and user subarray from
$4\times4$ UPA to $8\times8$ UPA. Since both the BS and users are
in the near-field of RIS, the locations of the center of the BS and
the RIS are generated randomly and are also shown in Table III. The
distance between the BS and the RIS is 1 m, while the distances between
the RIS and the users are 1 m, 0.5 m, 1.1 m and 0.8 m, respectively.
The simulation results are shown in Fig. \ref{fig:4 Achievable-rate-(bps/Hz)-1}
(a) and (b). Observe that our proposed scheme can also achieve much
higher spectral efficiency than the original system operating without
RIS. Furthermore, the spectral efficiency of the sensor-based PBA
scheme is higher than that of the scheme without PBA. Furthermore,
the spectral efficiency in the near-field scenario is much higher
than in the far-field scenario, as the path-loss is more severe in
the far-field case, which indicates that for better system performance,
we can use more RIS elements to create near-field communication for
the THz signals.

\section{Conclusions}

A novel hybrid 3D beamforming and sensor-based channel estimation
scheme is proposed in this paper for RIS-aided THz MIMO systems. The
user location obtained by UWB sensors is employed for RIS-BF and user
RBF. A PBA is also proposed for improving the beam alignment accuracy.
Compared to the benchmarks, our proposed scheme has lower complexity
and search time. We have also derived the conditions of channel orthogonality
for the RIS-aided THz channel to achieve high-integrity spatial multiplexing.
Moreover, the closed-form expressions of the near-field and far-field
path-loss of the RIS-aided THz channel are derived. Our simulation
results show that the proposed scheme accurately estimates the RIS-aided
THz channel and the spectral efficiency with RIS is much higher than
that without RIS. Both the near-field and far-field scenarios demonstrate
the benefits of our proposed scheme. As our future work, other forms
of ISAC will be explored for RIS-aided THz MIMO systems.

\section{Appendix A\protect \\
Proof of Theorem 1}
\begin{IEEEproof}
Let $\left(x,y\right)$ and $\left(\hat{x},\hat{y}\right)$ be the
coordinates of two arbitrary BS transmit subarrays and $\left(u,v\right)$
be arbitrary coordinates of a sub-RIS. The effective distance from
the centers of the BS subarray $\left(x,y\right)$ and the sub-RIS
$\left(u,v\right)$ is

\begin{equation}
d_{e}=\sqrt{d_{1}^{2}+\alpha^{2}\left[\left(u-x\right)^{2}+\left(v-y\right)^{2}\right]}.
\end{equation}
For $d_{1}\gg\alpha$, according to the binomial approximation, we
have 
\begin{equation}
d_{e}\approx d_{1}+\frac{\alpha^{2}\left[\left(u-x\right)^{2}+\left(v-y\right)^{2}\right]}{2d_{1}}.
\end{equation}

So we can readily derive the inner product between the corresponding
channel columns as
\begin{equation}
\begin{aligned} & \left\langle \mathbf{G}_{x,y},\mathbf{G}_{\hat{x},\hat{y}}\right\rangle \\
= & \left(\frac{c}{4\pi fd_{1}}\right)^{2}e^{-\mathcal{\mu}(f)d_{1}}\\
 & \times\sum_{u=0}^{N-1}\sum_{v=0}^{N-1}e^{k\frac{\pi f}{cd_{1}}\alpha^{2}\left[\left(u-x\right)^{2}+\left(v-y\right)^{2}-\left(u-\hat{x}\right)^{2}-\left(v-\hat{y}\right)^{2}\right]}\\
 & \cdot\mathbf{a}_{sa,x,y}\left(\delta_{x,y},\kappa_{x,y}\right)\mathbf{a}_{t,x,y}^{H}\left(\psi_{x,y},\sigma_{x,y}\right)\\
 & \mathbf{\cdot a}_{sa,k}\left(\delta_{\hat{x},\hat{y}},\kappa_{\hat{x},\hat{y}}\right)\mathbf{a}_{t,\hat{x},\hat{y}}^{H}\left(\psi_{\hat{x},\hat{y}},\sigma_{\hat{x},\hat{y}}\right)\\
= & \left(\frac{c}{4\pi fd_{1}}\right)^{2}e^{-\mathcal{\mu}(f)d_{1}}\\
 & \times\sum_{u=0}^{N-1}e^{k\frac{2\pi f}{cd_{1}}\alpha^{2}u(\hat{x}-x)}\sum_{v=0}^{N-1}e^{k\frac{2\pi f}{cd_{1}}\alpha^{2}v(\hat{y}-y)}\\
 & \cdot\mathbf{a}_{sa,x,y}\left(\delta_{x,y},\kappa_{x,y}\right)\mathbf{a}_{t,x,y}^{H}\left(\psi_{x,y},\sigma_{x,y}\right)\\
 & \cdot\mathbf{a}_{sa,k}\left(\delta_{\hat{x},\hat{y}},\kappa_{\hat{x},\hat{y}}\right)\mathbf{a}_{t,\hat{x},\hat{y}}^{H}\left(\psi_{\hat{x},\hat{y}},\sigma_{\hat{x},\hat{y}}\right).
\end{aligned}
\end{equation}
The channels are orthogonal when $\left\langle \mathbf{G}_{x,y},\mathbf{G}_{\hat{x},\hat{y}}\right\rangle =0$.
Therefore, without imposing any restrictions on the steering vectors,
we can obtain the following optimal value, 
\begin{equation}
\alpha_{op}=\sqrt{q\frac{d_{1}c}{Nf}}=\sqrt{q\frac{d_{1}\lambda}{N}}\label{eq:opt}
\end{equation}
for integer values of $q$. 
\end{IEEEproof}
\bibliographystyle{IEEEtran}
\bibliography{IEEEabrv,7D__Australia_thz_meeting_JCS________________1117________________1117}

\end{document}